\documentclass[onecolumn,final]{svjour2}
\usepackage{amstext}
\usepackage{amssymb}
\usepackage{graphicx}
\usepackage{amsmath}
\usepackage{color}

\newcommand{\Ip}{I_{\rm pred}}

\makeatletter
\@ifundefined{textcolor}{}
{%
 \definecolor{BLACK}{gray}{0}
 \definecolor{WHITE}{gray}{1}
 \definecolor{RED}{rgb}{1,0,0}
 \definecolor{GREEN}{rgb}{0,1,0}
 \definecolor{BLUE}{rgb}{0,0,1}
 \definecolor{CYAN}{cmyk}{1,0,0,0}
 \definecolor{MAGENTA}{cmyk}{0,1,0,0}
 \definecolor{YELLOW}{cmyk}{0,0,1,0}
 }

\makeatother
\renewcommand{\makeheadbox}{}

\usepackage[english]{babel}

\begin{document}

\title{Predictive information in a nonequilibrium critical model}

\author{Martin Tchernookov \and Ilya Nemenman}

\institute{Martin Tchernookov, \email{mtchern@.emory.edu} \at
  Department of Physics, Emory University, Atlanta, GA 30322, USA \and
  Ilya Nemenman, \email{ilya.nemenman@emory.edu} \at Departments of
  Physics and Biology and Computational and Life Sciences Initiative,
  Emory University, Atlanta, GA 30322, USA} \date{\today}
\maketitle
\keywords{phase transitions\and information theory\and subextensive scaling}
\PACS{05.70.Fh, 05.10.Gg, 89.70.-a, 89.75.-k}

\begin{abstract}
  We propose predictive information, that is information between a
  long past of duration $T$ and the entire infinitely long future of a
  time series, as a {\em universal order parameter} to study phase
  transitions in physical systems. It can be used, in particular, to
  study nonequlibrium transitions and other exotic transitions, where
  a simpler order parameter cannot be identifies using traditional
  symmetry arguments. As an example, we calculate predictive
  information for a stochastic nonequilibrium dynamics problem that
  forms an absorbing state under a continuous change of a
  parameter. The information at the transition point diverges as
  $\propto \log T$, and a smooth crossover to $\propto T^0$ away from
  the transition is observed.
\end{abstract}

\maketitle

\section{Introduction}

The theory of critical phenomena and the emergent notion of
universality was one of the singular developments of physics in the
twentieth century. With a known order parameter and symmetries of the
problem, calculation of long-range, measurable behaviors of
equilibrium physical quantities becomes a rather straightforward
task. The success has turned out to be hard to replicate for
non-equilibrium systems and systems where symmetry properties are
similar in the phases on both sides of the transition
\cite{Brazhkin:2012}. Here it is often unclear which quantity can
serve as a good order parameter, and the developed theoretical
machinery does not apply.  Where progress has been made, order
parameters have been very specific, making it difficult to identify
universal properties. For example, in reaction-diffusion problems with
absorption, one commonly uses linear superposition of particle
concentrations as order parameters \cite{Tch:2010,vanWijland:1998},
while particle current is a better choice for jamming problems
\cite{Garrahan:2007}. Further, the order parameters often have
nontrivial relations to easily observable quantities. For example,
phase transitions in some systems with dynamic heterogeneities often
must be described with four-point correlation functions of particle
densities \cite{Biroli:2007}, or a multitude of correlation functions
\cite{Binder:1990,LeDoussal:2004ua}. Similarly, dynamical phase
transition require one to study the space of trajectories instead of
the state space \cite{Lecomte:2007}.

Whatever the choice, the order parameter is a statistics averaged over
a distribution of microscopic states. A continuous or discontinuous
change in its value at a transition indicates a similar change in the
underlying probability distribution. Therefore, it is natural to shift
attention to the distribution itself, and, specific to nonequilibrium
systems, to how it converges to the steady state.

Intuitively, different phases (often with different symmetries)
manifest themselves by changes in our ability to use local
experimental measurements for long-range predictions. For example,
nonzero magnetization in an Ising magnet allows us to predict with
some certainty orientation of far away spins based on the value of the
spin at the origin. Similarly, different crystalline phases of solids
have different density autocorrelation functions, and hence existence
of an atom at the origin translates into different predictions about
the presence of an atom a certain distance away. Then instead of a
specific statistics characterizing the predictability, namely the
order parameter, it might be useful to study one's ability to use
local measurements to predict states of the rest of the system {\em
  directly}.

This prediction ability is naturally quantified using the language of
Shannon's information theory \cite{Shannon:1998ti}. In previous work,
we have termed it the {\em predictive information}
\cite{Bialek:2001wv,Bialek:2001wz}. Briefly, in information theory,
the total uncertainty in a system specified by a state $\vec{x}\in X$,
$\dim \vec{x}=N$, is measured by the (differential) entropy,
\begin{equation}
  S[X]=-\int d^Nx P(\vec{x})\log_2P(\vec{x}).
\end{equation}
Then observing a state of another variable $\vec{y}\in Y$, $\dim y=M$,
may reduce the uncertainty about $\vec{x}$, and hence provide the
information about it
\begin{eqnarray}
  I[X;Y]&=&S[X]-\langle S[X|Y]\rangle_Y= \int d^Nx\,d^My\,P(\vec{x},\vec{y})
  \log \frac{P(\vec{x},\vec{y})}{P(\vec{x})P(\vec{y})}\nonumber\\
&=&  \left<
  \log \frac{P(\vec{x},\vec{y})}{P(\vec{x})P(\vec{y})}\right>_{X,Y}
=I[Y;X].
\label{eq:pred_Defined}
\end{eqnarray}
Importantly, $I[X;Y]$ depends on the entire probability distribution
$P(\vec{x},\vec{y})$, but not just on its specific statistics, and it
is zero iff $X$ and $Y$ are statistically independent. 

One can consider $X$ and $Y$ to be states of a physical process, such
that $X$ are the measured quantities, and $Y$ are the quantities that
one wants to predict \cite{Bialek:2001wv}. For example, $X$ can be the
state of spins on one segment of an Ising chain, and $Y$ be the state
of spins far away. Similarly, for time series and for nonequilibrium
processes, $X$ can be the past of the process of duration $N$, and $Y$
part of its future of duration $M$.  Then the information becomes the
{\em predictive information}:
\begin{equation}
I_{\rm pred}(N,M)=I[X;Y].
\label{eq:Ipred}
\end{equation}
Since the quantification of the intrinsic state of the system should
not depend on which specific set of variables $Y$ one wants to
predict, it makes sense to define predictive information as
\begin{equation}
  I_{\rm pred}[X]\equiv I_{\rm pred}(N)=\lim_{M\to\infty} I(N,M).
\label{eq:IN}
\end{equation}
That is, one quantifies how much information the local observations
$X$ provide about an entire, infinitely large physical system.

Predictive information is subextensive, $\lim_{N\to\infty} \Ip(N) /
N=0$ \cite{Bialek:2001wv}. It tends to a handful of universal
behaviors for large systems, $N\to\infty$, intuitively correlating
with the complexity of the underlying physical process. In particular,
$\lim_{N\to\infty}\Ip(N)={\rm const}$ indicates an easily predictable
deterministic, or a short correlation length probabilistic dynamics
(``simple'' long range prediction can be perfect, or it is impossible,
respectively). Further, $\lim_{N\to\infty}\Ip(N)\propto \log N$ is
indicative of a second order equilibrium phase transition (power-law
decaying correlations allow for complex, multiscale, partially
predictable patterns over very long distances). Finally, $
\lim_{N\to\infty}\Ip(N)\propto N^\alpha$, $\alpha<1$ may correspond to
more exotic phase transitions with infinite-dimensional order
parameters, but this case is not well understood.

The dependence of $\Ip$ on the full underlying probability
distribution and the relation to phase transitions make it natural to
explore $\Ip$ as a ``universal order parameter'', also useable in the
nonequlibrium context. However, we are not aware of calculations of
predictive information for nonstationary processes, where $P(\vec{x})$
is explicitly or implicitly time dependent. Further, even for
equilibrium systems, the transition between $\Ip={\rm const}$ and
$\Ip\propto \log N$ in the vicinity of a phase transition has not been
studied.

In this paper, we study predictive information in a context of a
simple nonequilibrium, continuous-time Markov process, which ages and
develops an absorbing state at a certain critical value of a
parameter. This process can be viewed as a toy model, which is likely
to possess some features of more complex systems. We calculate the
expression for predictive information at the critical point and, for
the first time for any system, near the critical point.  The
calculation reveals the need to modify the definition,
Eq.~(\ref{eq:IN}), to remove an ultraviolet divergence emerging due to
the continuous-time nature of the process. Similar modifications will
likely allow extension of predictive information methodology to
multidimensional systems.  We demonstrate explicitly the logarithmic
divergence of $\Ip$ at the transition, and we show that the divergent
term in the information is insensitive to temporally local, invertible
transformations of the state space. This makes predictive information,
and specifically its divergent term, a great candidate to characterize
nonequlibrium phase transitions.

\section{The model}
We consider a Markovian system governed by the following Langevin
equation:
\begin{align}
\label{eqn:lang}
&\partial_tx(t)=-x(x^2+\tau)+\sqrt{2}\sigma |x|^{\alpha/2}\eta ,\\
&x(t=0)=x_0, \;\mbox {sampled from } P(x_0)\equiv P_0,
\end{align}
where $\langle\eta(t)\eta(t')\rangle=\delta(t-t')$. We will treat this
equation in the Ito sense. Without the noise term, $x$ relaxes from
the initial value $x_0$ to either 0 or $\pm \sqrt{\tau}$, depending on
if $\tau>0$. The transition happens at $\tau=0$. For large noise near
$x=0$ (that is, small $\alpha$), $x$ gets kicked out from $x\approx0$
region, and the system equilibrates. For small noise (large $\alpha$),
a near-deterministic relaxation to the absorbing state at $x=0$
persists. This is probably the simplest example of nonequilibrium,
stochastic relaxation dynamics, and it is a natural starting point for
the analysis.

We note that we can view this equation as describing dynamics of
magnetization, $x$, along a line normal to a boundary of an Ising
ferromagnet in some number of spatial dimensions. The coordinate is
$t=0$ at the boundary, and increases into the bulk. The deterministic
cubic dynamics in Eq.~(\ref{eqn:lang}) is the usual coarse-grained
model of such ferromagnet. In such a model, the variance of the noise
increases with $x$, and $\alpha$ would depend on the overall
dimensionality of the problem.

To calculate predictive information, Eq.~(\ref{eq:Ipred}), we
discretize the time $t$, $t_n=n\Delta t$, and $x_n=x(t_n)$. We choose
$\Delta t\to 0$, and yet $N\Delta t=T_{\rm p}\to\infty$, and $M\Delta
t=T_{\rm f}\to\infty$, where p and f stand for {\em past} and {\em
  future}, respectively.  Then Eq.~(\ref{eqn:lang}) is equivalent to
the following Markovian dynamics:
\begin{multline}
P(x_{n+1}|x_0,x_1,...,x_n)=P(x_{n+1}|x_n)\\=
\frac{1}{\sqrt{4\pi\Delta t}\sigma x^{\alpha/2}}\exp\left\{-\frac{\left[x_{n+1}-\left(x_n-x_n(x_n^2+\tau)\Delta
        t\right)\right]^2}
{4\sigma^2 |x_n|^\alpha\Delta t} \right\}.
\end{multline}
To simplify the notation, we define
\begin{align}
  &P_{n|n-1}\equiv P(x_n|x_{n-1}),\\
  &P_{n}\equiv P(x_n)=\int dx_{n-1}P(x_{n-1})P(x_{n}|x_{n-1}).
\end{align}
Then:
\begin{multline}
\label{eqn:minminfo}
I_{\rm pred}(N,M)= 
\left\langle \log_2\frac{ P_0\prod_{n=1}^{N+M-1}P_{n|n-1}} {
    P_0\prod_{n=1}^{N-1}P_{n|n-1} P_N
    \prod_{m=N+1}^{N+M-1}P_{m|m-1}}\right\rangle \\
=\left\langle\log_2 \frac{P_{N|N-1}}{P_N}\right\rangle=I[x_N;x_{N-1}].
\end{multline}

Not surprisingly for a Markovian process, predictive information
is the mutual information between two successive measurements and does
not depend on the length of the future sequence, $M$, so that the
limit, Eq.~(\ref{eq:IN}), is trivial. However, the information can
depend on $N$ since the system is not stationary, and not
time-translation invariant. Specifically, for small noise, each
subsequent $x$ is more narrowly distributed. This allows the
information to increase unboundedly with $N$, unlike in typical
finite-dimensional Markov processes with constant transition
probabilities, where $I_{\rm pred}$ is always finite
\cite{Bialek:2001wv}. These considerations also point out that one
must take the sequence on $N$ observations starting from exactly the
same time when calculating the averages.

Since $x(t)$ is continuous, $x_{N}\to x_{N-1}$ as $\Delta t\to 0$. The
state of the process at the next time step becomes exactly known, and
predictive information diverges. However, this is a superficial
ultraviolet divergence, while we are interested in studying the
infrared behavior. Interestingly, this interfacial effect has been the
primary reason behind the inability to apply predictive information
ideas to systems in more than one dimension, where the size of the
interface diverges with the system size. This makes it difficult to
disambiguate divergences in predictive information coming from
long-range prediction from those produced by short range interfacial
effects.

We thus need to introduce the cutoff scale into the system, at which
predictive information is computed, similarly to how one does this in
the renormalization group theory. For this, we redefine predictive
information as mutual information between the past of duration $T_{\rm
  p}=N\Delta t$ and the future of duration $T_{\rm f}=M\Delta T$,
separated by a ``scale'' gap of duration $T_{\rm s}=L\Delta T$, which
remains finite as $\Delta T\to0$. That is
\begin{multline}
\label{eqn:geninfo}
I_{\rm pred}(N,M|L)= \\
\left\langle \log_2\frac{ P_0 \prod_{n=1}^{N-1}P_{n|n-1}P_{N+L|N-1}\prod_{m=N+L+1}^{N+L+M-1} P_{m|m-1}}
  {P_0 \prod_{n=1}^{N-1}P_{n|n-1}P_{N+L}\prod_{m=N+L+1}^{N+L+M-1} P_{m|m-1}}\right\rangle \\ =\left\langle\log_2
  \frac{P_{N+L|N-1}}{P_{N+L}}\right\rangle=I[x_{N+L};x_{N-1}].
\end{multline}
Here
\begin{equation}
P_{N+L|N-1}=\int \prod_{n=N}^{N+L-1}dx_n\prod_{m=N}^{N+L} P_{m|m-1}.
\end{equation}

\section{Invariance of predictive information}
From Eq.~(\ref{eqn:geninfo}), it is clear that predictive
information is invariant under reparameterization of $x$. This is a
desired property for any potential universal order parameter. Further,
any experimental device measuring $x(t)$ will act as a temporal
filter, so that the measured values will be convolutions of true $x$'s
at nearby time points. Thus it is also desirable for the
nonequilibrium order parameter to be invariant to temporally local
invertible transformations of data \cite{Bialek:2001wv}. Does the
predictive information obey this property?

The filter, represented by ${\mathcal F}$, maps the sequences of true
states of the system $\{x\}$ into measured data $\{\chi\}$.  We
require that the filter does not inject additional information into
the dynamics. This means that the extraneous parameters of the mapping
$\mathcal{F}$ must be known. In a real-life experiment, this means
that we would like to be able to separate the behavior of the observed
system from any artifacts associated with the experimental setup.

In general terms, such filter can be represented by a convolution
kernel $\mathcal{L}(t-t')$. Since a convolution mixes the past and the
future, the measured data $\{\chi\}$ is no longer Markovian. We
require that the so-introduced statistical dependences are short
lived, i.\ e.\, the kernel $\mathcal{L}(t-t')$ is of compact support
or decreases with time exponentially or faster. This is our definition
of temporal locality.

Convolutions are reductions in rank and therefore (potentially)
invertible only for infinitely long data sequences.  Therefore, we can
define invertibility only in the $t\to\infty$ limit. To this end, let
$\mathfrak{V}=\bigotimes_n \mathbb{R}^n$ be the space of all
temporally discretized, finite length trajectories, that is the space
of all $n$-tuples of $x$, $n<\infty$. Let $\mathcal{F}:
\mathfrak{V}\to\mathfrak{V}$ be a function such that $\mathcal{F}(
\mathbb{R}^{N+\nu})\subset\mathbb{R}^N$. That is, a sequence of $N$
data points is defined from $N+\nu$ points through some filtering
procedure. We consider this mapping to be invertible if the
Radon-Nikodym derivative over the set $\mathcal{F}^{-1}\left(\mathbf
  x\in \mathbb{R}^N\right)$ converges to a delta function for
$N\to\infty$. More specifically, the probability of observing a
trajectory $\{\chi_i\}_{i=1}^N$ is given by
\begin{multline}
\label{eqn:probchi}
P(\{\chi_j\}_{i=j}^N)= \int\,d^{N+\nu}xP(\{x_j\}_{j=-\nu}^N) 
\prod_{j=1}^N \delta\left(\chi_j-\sum_k \mathcal{L}(j-k)x_k  \right) \\ 
=\int\,d^{N+\nu}x\,d^N\lambda\times \\ \times \exp\left[
-i\sum_{j=1}^N\lambda_j \left(\chi_j-\sum_k \mathcal{L}(j-k)x_k  \right)
+\ln P(\{x_j\}_{j=-\nu}^N) \right].
\end{multline}
Thus invertibility requires that the Hessian matrix of the exponent in
this equation diverges, defining a dominant stationary solution of the
corresponding ``action''. With this requirement, $\{\chi_i\}$ are
simply reparameterizations of $\{x_i\}$, and predictive
information is invariant under the change. While this requirement is
very general, we suspect that, in practice, it will be equivalent to
the asymptotic properties of trajectory-averaged quantities, for which
there are already well established results \cite{jones:2004}. We leave
exploration of these conditions to future work.

\section{Solving the model}
To calculate predictive information in the model, we first
calculate the Green's functions (the marginal and the conditional
distributions) of Eq.~(\ref{eqn:lang}). For this, we write the
Fokker-Planck equation corresponding to the Langevin dynamics
\begin{equation}
\label{eqn:FPgen}
\partial_tp(x,t)=\partial_x\left[x(x^2+\tau)p(x,t)+\sigma^2 \partial_x\left(|x|^{\alpha} p\left(x,t\right)\right)\right].
\end{equation}
This equation immediately confirms our earlier statement that
$p(x,t)=\delta(x)$ is a stationary state, stability of which depends
on the strength of the noise, which in turn is controlled by
$\alpha$. As a result, the equation can develop a singularity near
$x=0$.  Fortunately, the probability current at $x=0$ is zero. Thus
for $x_0>0$, we can consider $x(t)>0$ for any $t$.  Further, we seek
the solution for $\tau>0$, hoping further to analytically continue to
the entire real axis of $\tau$. With these caveats, we make the
following simplifying transformations:
\begin{align}
&\bar{\tau}\equiv\frac{\beta^2}{\sigma^2}\hat\tau=\beta\tau/\sigma^2,\\
&\hat{t}=t\tau/\beta,\\
&\hat y\equiv y\hat{\tau}^{1/2}=x^{-1/\beta}\hat{\tau}^{1/2},\\
&f=y^{-\beta\alpha}p\left(x\left(y\right),t\right),\\
&\beta=2/(\alpha-2),\\
&n=2(\alpha-1)/(\alpha-2).\label{eq:n}
\end{align} 
Then Eq.~(\ref{eqn:FPgen}) becomes
\begin{equation}
\label{eqn:al5diff}
\hat{y}^{n-1}\partial_{\hat{t}} f=-\partial_{\hat{y}}\left[\left(
  \hat{y}^n+\frac{\beta\bar{\tau}^{(n-3)}}{\sigma^2}\hat{y}^{4-n}\right)f\right]+
\partial_{\hat{y}}\left(\hat{y}^{n-1}\partial_{\hat{y}} f\right) . 
\end{equation}
The initial condition should obey $p(\hat y=0,t)=p(\hat y\to\infty,t)=0$. The
former condition is a result of the inverse relationship between $x$
and $\hat y$, while the latter is due to $x=0$ being the absorbing state.

It is important to discuss the allowed values of $\alpha$ at this
point. From Eq.~(\ref{eq:n}), $n$ becomes divergent at $\alpha=2$.
This corresponds to a large noise, which hides the phase transition.
On the other hand, for large $\alpha$, the noise is negligible, and
the system is in an effectively deterministic regime. This happens at
$n\le 3$, where the second term in
Eq.~(\ref{eqn:al5diff}) is suppressed as $\bar\tau\to 0$. Thus we are
interested in $3< n<\infty$, which corresponds to $2<\alpha<4$. In
this regime, the $\bar{\tau}$ term in Eq.~(\ref{eqn:al5diff}) is
negligibly small, and can be dropped.

With this, we notice that Eq.~(\ref{eqn:al5diff}) is the radial part
of the diffusion equation in $n$ dimensions. Thus our strategy is to
solve it first for $n$ integer, hoping to analytically continue to all
$n$ later on.  Assuming an integer $n$, we rewrite
Eq.~(\ref{eqn:al5diff}):
\begin{equation}
\label{eqn:al6diff}
\partial_{\hat{t}} f=- nf -{\hat y}\partial_{\hat y} f
 +  \frac{1}{{\hat y}^{n-1}}\partial_{\hat y}\left({\hat y}^{n-1}\partial_{\hat y} f\right).
\end{equation}
Therefore, $f(\hat y)$ is the radially symmetric part of the solution of
the following equation
\begin{equation}
\label{eqn:al7diff}
\partial_{\hat{t}} f=-n f - 
\mathbf{\hat y}\cdot\mathbf{\nabla} f + \nabla^2 f.
\end{equation}
We solve this equation in Appendix A, resulting in:
\begin{multline}
\label{eqn:ugrn}
G(t,y,z)= C(n)z^{n-1}\left(\frac{\hat\tau}{2\pi (e^{2 \hat\tau
      t}-1)}\right)^{n/2}\times\\
\int_{-1}^1 \,d \lambda \exp\left(-\frac{\hat\tau}{2(e^{2\hat\tau
      t}-1)}(y^2-2yze^{\hat\tau t}\lambda+z^2e^{2\hat\tau t})\right) K(\lambda),
\end{multline}
where $K(x)$ is a kernel, which, for integer $n$, is the Jacobian of
the $n$-dimensional change of variables from Cartesian to spherical
coordinates. We still need to determine it for non-integer
dimensions. For this, we substitute the expression of
Eq.~(\ref{eqn:ugrn}) in Eq.~(\ref{eqn:al6diff}) (for general $n$) and
find that it satisfies iff given by
\begin{equation}
\label{eqn:K}
\partial_{\lambda}^2[(1-\lambda^2)K(\lambda)]+(n-1)\partial_{\lambda}(\lambda K(\lambda))=0
\end{equation}
To guarantee regularity at $\lambda=\pm 1$ (and in analogy with the 
integer dimensional cases), we additionally impose the condition 
that $K(\pm 1)=0$, leading to the solution
\begin{equation}
K(\lambda)=(1-\lambda^2)^{\frac{n-3}{2}}.
\end{equation}

The normalization constant $C(n)$ can be determined from the
requirement that the integral over $y$ for a fixed $z$ is unity when
$t\to 0$.  In the case of an integer $n$, $C(n)$ is the area of the
unit sphere in $n-1$ dimensions. To verify this for any value $n$, we
need to perform the integration explicitly. To this end, it is
convenient to introduce $\Delta=[(e^{2\hat\tau t}-1)/\hat\tau]^{1/2}$,
and $z'=ze^{\hat\tau t}$. Then integrating  Eq.~(\ref{eqn:ugrn}), we get
\begin{multline}
\label{eqn:intgrn}
\int_0^{\infty}G(t,y,z)\,dy=C(n)z^{n-1}\left(\frac{1}{\sqrt{2\pi}
    \Delta}\right)\int_{0}^{\infty} dy \exp\left(-\frac{(y-z')^2}{2\Delta^2}\right)\\
\int_{-1}^1(\sqrt{2\pi})^{1-n}\Delta^{1-n}\exp\left(-\frac{yz'(1-\lambda)}{\Delta^2}\right) K(\lambda) \,d\lambda.
\end{multline}
We concentrate on the inner integral first. We perform the
substitution $\xi=yz'(1-\lambda)/\Delta^2$ which leads to
\begin{multline}
\label{eqn:int1grn}
\int_0^{2yz'/\Delta^2} (yz')^{-\frac{n-1}{2}}(\sqrt{2\pi})^{1-n}e^{-\xi}\left[\xi\left(2-\frac{\Delta^2 \xi}{yz'}\right)\right]^{\frac{n-3}{2}}\,d\xi\xrightarrow[\Delta\to\infty]{}\\
\frac{(yz)^{-\frac{n-1}{2}}}{2\pi^{(n-1)/2}}\int_{0}^{\infty}e^{-\xi}
\xi^{\frac{n-3}{2}}\,d\xi 
=\frac{1}{2\pi^{(n-1)/2}}(yz)^{-\frac{n-1}{2}}\Gamma\left(\frac{n-1}{2}\right).
\end{multline}
By dominated convergence, the limit is valid for any $y$ and all $n>
1$.  (The cases $3\ge n> 1$ follow from the fact that
$\xi(2-\Delta^2 \xi/yz')\ge \xi$ for $0<\xi\le
yz'/\Delta^2$, while the portion of the integral in Eq. \ref{eqn:int1grn}
between $yz'/\Delta^2<\xi\le 2yz'/\Delta^2$ converges to $0$
as $\Delta\to 0$).
Furthermore, since $yz'/\Delta^2$ controls the convergence in a 
monotonic fashion, the limit
is uniform on any semi-infinite interval not containing 0. Since the
convergence is dominated by a multiple of $(yz)^{-(n-1)/2}$,
particularly for the values of $y$ close to zero, we recognize the
outer integral in Eq.~(\ref{eqn:intgrn}) as a delta function.
Therefore, in order to bring the value of Eq.~(\ref{eqn:intgrn}) to
unity, we need that 
\begin{equation}
  C(n)=\frac{2\pi^{(n-1)/2}}{\Gamma((n-1)/2)},
\end{equation}
which is the area of the $n-1$ dimensional unit sphere when $n$ is
integer.

By reverting back to the original coordinate $x$, we can rewrite
Eq.~(\ref{eqn:ugrn}) and obtain the solution in these coordinates.
However, for the purposes of the next section, it is more convenient
to stay in the $y$ space instead. Notice that if we make the
substitutions $\tilde p=y^{-\alpha\beta/2}p$ in Eq.~(\ref{eqn:FPgen}),
we obtain
\begin{equation}
\label{eqn:normdiff}
\partial_t\tilde p=-\frac{1}{\beta}\partial_y\left((\hat\tau
  y+\frac{\alpha\sigma^2}{2}y^{-1}+y^{5-2n})\tilde p\right)+
\frac{\sigma^2}{\beta^2}\partial_y^2\tilde p .
\end{equation}
The advantage of $\tilde p$ over $f$ calculated earlier is that
$\tilde p$ is a probability distribution. We can immediately write its
Green's function from Eq.~(\ref{eqn:ugrn}) since $\tilde
p(t,y)=y^{n-1}f(t,y)$:
\begin{multline}
\label{eqn:fgrn}
\tilde G(t,y,z)=C(n)(y)^{n-1}\left(\frac{\hat\tau}{2\pi (e^{2 \hat\tau t}-1)}\right)^{n/2} \times\\ 
\times\int_{-1}^1 \,d \lambda
\exp\left(-\frac{\hat\tau}{2(e^{2\hat\tau t}-1)}(y^2-2yze^{\hat\tau
    t}\lambda+z^2e^{2\hat\tau t})\right) K(\lambda) .
\end{multline}
This is the main result of this section, which we will use in order
calculate predictive information for our model.  One can verify by
explicit substitution that the expression in Eq.~(\ref{eqn:fgrn})
satisfies the Fokker-Planck equation, Eq.~(\ref{eqn:al5diff}), and it
reduces to a delta function as $t\to 0$. Thus it represents the
conditional distribution of $y$ given $z$.

\section{Predictive information for the model}
\begin{figure}
\label{fig:1}
\includegraphics[width=0.75\textwidth]{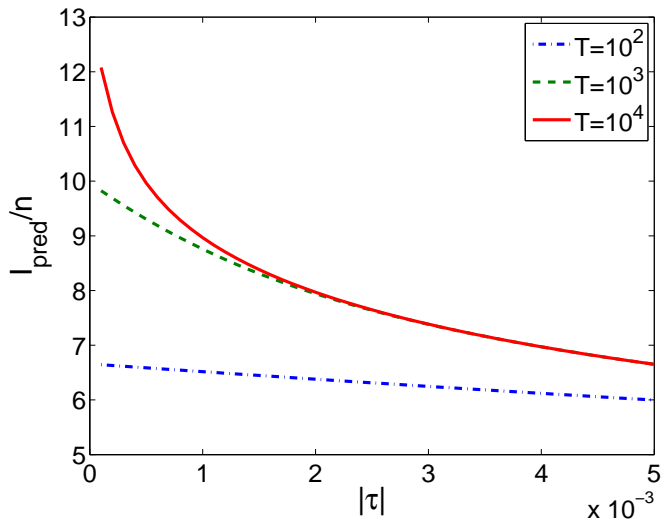}
\caption{A plot of $\Ip/n$ for different values of $\tau<0$ at
different times $T$ for a fixed $\tilde t=1$.}
\end{figure}

Predictive information is reparameterization invariant. Thus we
can calculate it for $y$ instead of $x$ and use the expression,
Eq.~(\ref{eqn:fgrn}), when applying the Eq.~(\ref{eqn:geninfo}) to our
model. Without loss of generality, we assume that the initial
condition is a delta function. Then the continuous form of
Eq.~(\ref{eqn:geninfo}) is
\begin{equation}
\label{eqn:geninfocont}
\Ip (t)=\left\langle 
\log_2\frac{\tilde G(\tilde t,y,z)}{\tilde G(t+\tilde t,y,w)}
\right\rangle,
\end{equation}
where $w$, $z$, and $y$ are the values of the observable at times $0$,
$t=(N-1)\Delta t$, and $T\equiv t+\tilde t=(N+L)\Delta t$
respectively, i.\ e., $w=x_0^{-1/\beta}$, $z=x_{N-1}^{-1/\beta}$, and
$y=x_{N+L}^{-1/\beta}$.  Equation (\ref{eqn:geninfocont}) involves an
integral with complex time and $\hat\tau$- dependences. In the
following, we would like to find the leading orders of these
dependences. Defining $\Delta(t)=[(e^{2\hat\tau t}-1)/\hat\tau]^{1/2}$
(cf.~Eq.~(\ref{eqn:ugrn})), it is also convenient to introduce
$\Xi(t;\lambda,y,w)=\exp[(y^2-2ywe^{\hat\tau t}\lambda+w^2e^{2\hat\tau
  t})/ (2\Delta(t)^2)]$, so that Eq.~(\ref{eqn:fgrn}) takes on the form
\begin{equation}
\label{eqn:fgrn1}
\tilde G(t,y,z)=C(n)\frac{(2\pi)^{-n/2}}{\Delta(t)^n}
\int_{-1}^1 \,d\lambda\, K(x)\, \Xi (t;\lambda,y,z).
\end{equation}
Then Eq.~(\ref{eqn:geninfocont}) becomes
\begin{multline}
\label{eqn:geninfocont1}
\Ip(t)=n\log_2\frac{\Delta(T)}{\Delta(\tilde t)}+
\left\langle\log_2
\int_{-1}^1 \,d\lambda K(\lambda) \Xi (\tilde t;\lambda,y,z)
\right\rangle- \\
\left\langle\log_2
\int_{-1}^1 \,d\lambda K(\lambda) \Xi (T;\lambda,z,w)
\right\rangle.
\end{multline}
In Appendix C, we show that the last two terms in
Eq.~(\ref{eqn:geninfocont1}) are asymptotically constant when
$T\to\infty$ if $t$ is large and $\hat\tau$ is small. Therefore, to
the leading order, predictive information is
\begin{equation}
\label{eqn:infofinal}
\Ip(t)\approx n\log_2\frac{\Delta(T)}{\Delta(\tilde t)}= n\log_2\frac{\exp[2\hat\tau(t+\tilde t)]-1}{\exp(2\hat\tau \tilde t)-1}.
\end{equation}
At the critical point, when the absorbing state is just starting to
emerge, $\hat\tau\to 0$, this expression reduces to
\begin{equation}
\label{eqn:infofinal0}
\Ip(t)\approx n\log_2\frac{t+\tilde t}{\tilde t}.
\end{equation}
This logarithmic growth with the system size $t$ has been anticipated
for a critical point in Ref.~\cite{Bialek:2001wv}, but has not been
calculated before for any nonequilibrium stochastic dynamical
system.  A plot of Eq.~(\ref{eqn:infofinal}) is given for different
parameter values in Fig.~\ref{fig:1}.

Notice that the prefactor $n=2(\alpha-1)/(\alpha -2)$ increases with
the effect of the noise, which corresponds to more of partially
predictable variability in the dynamics, and hence to an intuitively
higher complexity. Further, as $\alpha\to2$, or $n\to\infty$, the
leading term in predictive information becomes extensive, and
 hence it would cancel out in the difference of entropies in
Eq.~(\ref{eq:pred_Defined}), leading to $I_{\rm pred}(t)={\rm const}$.
Equation~(\ref{eqn:infofinal}) also allows calculation of the
asymptotic away from the phase transition. For large negative
$\hat{\tau}$, $I_{\rm pred}(t)={\rm const}$. For large positive $\tau$,
$I_{\rm pred}(t)\propto t$, since perfect prediction is possible in the
absorbing state. Hence it cancels out as well, leading to the constant
limit, and indicating the absence of the phase transition. These
results illustrate that divergence of predictive information correctly
captures the existence of the phase transition (emergence of the
absorbing state) at $\tau\to0$.

\section{Discussion}
Predictive information was introduced in Ref.~\cite{Bialek:2001wv} as
information between the past and the future of a time series, or
between left and right parts of a physical system. It was argued, in
particular, that the behavior of predictive information as the system
size grows can signal existence of a phase transition. As an example,
Ref.~\cite{Bialek:2001wv} calculated the information numerically for
an equilibrium long-range one-dimensional Ising magnet. In the
current work, we argue that predictive information can be used as a
universal order parameter in more complicated scenarios, such as in
nonequilibrium contexts, where traditional symmetry arguments fail to
identify low-order correlation functions that can serve this role. For
the first time, we calculate predictive information for a
nonequilibrium Markov process, which exhibits a phase transition at
certain values of parameters. Divergence of predictive information
correctly captures this phase transition. In addition to results {\em at}
and {\em far away} from the critical point, our calculations reveal how
predictive information behaves {\em near} a phase transition,
exhibiting a smooth crossover from an asymptotically constant to an
asymptotically divergent regime.  To our knowledge, this has not been
calculated before, either for equilibrium or for nonequlibrium
systems.

One important technical difference between this work and the previous
ones is the introduction of an additional ``renormalization'' scale,
$L$ or $\tilde{t}$, in the definition of predictive information,
so that the information is calculated between the past and the future
that are separated by a finite distance. This removed the ultraviolet
divergences associated with information at the interface between the
past and the future of a trajectory. While this modification was
precipitated by the continuous time/space nature of the stochastic
process, we believe that it will solve additionally difficulties with
application of predictive information ideas to systems with more than
one dimension. Indeed, there the main problem is that the interface
between two parts of a system diverges with the system size, and hence
the interfacial contribution to predictive information diverges
even away from a critical point. This will not happen if direct
interfaces are eliminated.

In summary, in this paper, we provide the first example of a direct
analytical calculation of predictive information for a nonequilibrium
stochastic process. This example argues further for using predictive
information as a universal order parameter for studying phase
transitions.

\begin{acknowledgements}
  This work has been supported in part by a James S.\ McDonnell
  Foundation Complex Systems Grant No.~220020321. We would like to
  thank HGE Hentschel for stimulating discussions.
\end{acknowledgements} 

\appendix

\section{Calculating the Green's function}
Green's function of Eq.~(\ref{eqn:al7diff}) is found easier in the
Cartesian coordinates, and the radial component can be extracted
afterwards. Thus we look for the Green's function of the form
\begin{equation}
\label{eqn:grn}
G(\hat{t};\vec{\hat y},\vec{\hat z})=\prod_i^n G_1(t;\hat y_i,\hat z_i)
\end{equation}
where $G_1(\hat{t};\hat y_i,\hat z_i)$ is the one dimensional Green's function,
satisfying
\begin{equation}
\label{eqn:al8diff}
\partial_{\hat{t}} G_1=- G_1 - {\hat y}\partial_{\hat y}G_1 + \partial_{\hat y}^2 G_1 
+\delta(\hat{t},\hat y- \hat z). 
\end{equation}
 
To solve Eq.~(\ref{eqn:al8diff}), it is convenient to consider $\tilde
G_1=e^{ \hat t}G_1$, where $\tilde G_1$ satisfies
\begin{equation}
\label{eqn:al1grn}
\partial_{\hat{t}} \tilde G_1(\hat{t};\hat y,\hat z) -\partial_{\hat y}^2 \tilde G_1(\hat{t};\hat y,\hat z)+
\hat y  \partial_{\hat y} \tilde G_1(\hat{t};\hat y,\hat z)=\delta(\hat{t},\hat y-\hat z).
\end{equation}
As usual, we transform into Fourier space:
\begin{equation}
\label{eqn:al2grn}
i\omega\tilde G_1 + k^2 \tilde G_1 -\partial_k(k\tilde G_1)=e^{-ik\hat z}.
\end{equation}
If we use the integral multiplier 
\begin{equation}
\mu=\exp\left(-(i\omega \ln k+k^2/2)\right),
\end{equation}
we obtain the following simplified form of Eq.~(\ref{eqn:al2grn})
\begin{equation}
-\partial_k (k\mu \tilde G_1)=\mu e^{-ik\hat z}.
\end{equation}
Since we are looking for a smooth solution, we expect $\tilde G=0$ as
$k\to\infty$. Therefore, the correct solution of the above equation is
in the form 
\begin{equation}
\tilde G_1(\omega,k,\hat z)=k^{-1}\mu^{-1}\int_k^{\infty} e^{-ik'\hat z}e^{-(iw\ln k' +k'^2/2)} \,dk'.
\end{equation}
Inverting back to the time coordinate, we obtain 
\begin{equation}
\tilde G_1(\hat t,k,\hat z)=e^{k^2/2} k^{-1}\int_k^{\infty} e^{-ik'\hat z}e^{-k'^2/2} \delta(\hat t-\ln k'+\ln k)\,dk'.
\end{equation}
Now performing the delta function integration, we are left with
\begin{equation}
\tilde G_1(\hat t,k,\hat z)=e^{k^2/2}e^{\hat t} e^{-ike^{\hat t} \hat z-k^2e^{2\hat t}/2}.
\end{equation}
This is simply a Gaussian function, and the transformation back to the
$\hat y$ coordinate leaves us with
\begin{equation}
\label{eqn:al3grn}
G_1(\hat t,\hat y,\hat z)=e^{-\hat t} \tilde G_1(\hat t,\hat y,\hat z)=\left[2\pi (e^{2 \hat t} -1)\right]^{-1/2} \exp\left(-\frac{1}{2}
  \frac{(\hat y-e^{\hat t} \hat z)^2}{e^{2 \hat t}-1}\right).
\end{equation}
We would like to extract the full dependence of the above solution on
$\hat\tau$. For normalization purposes, it is also convenient to
multiply by $\hat\tau^{1/2}$. Thus rescaling back to the $t$ and $y$
coordinates results in
\begin{equation}
\label{eqn:al4grn}
G_1(t,y,z)= \left(\frac{\hat\tau}{2\pi(e^{2\hat\tau t}-1)}\right)^{1/2}\exp\left(-\frac{\hat\tau}{2}
  \frac{(y-e^{\hat\tau t} z)^2}{e^{2 \hat\tau t}-1}\right).
\end{equation}

This finally results in an expression for the Green's function of
Eq.~(\ref{eqn:al8diff}), which in turn gives the Green's function of
Eq.~(\ref{eqn:al7diff}) in Cartesian coordinates. Now, to obtain the
solution of Eq.~(\ref{eqn:al6diff}), we need to revert back to
spherical coordinates. The resulting expression when $n$ is integer
suggests that we look for $G(t,y,z)$ in the following form
\begin{multline}
G(t,y,z)= C(n)z^{n-1}\left(\frac{\hat\tau}{2\pi (e^{2 \hat\tau
      t}-1)}\right)^{n/2}\times\\
\int_{-1}^1 \,d \lambda \exp\left(-\frac{\hat\tau}{2(e^{2\hat\tau
      t}-1)}(y^2-2yze^{\hat\tau t}\lambda+z^2e^{2\hat\tau t})\right) K(\lambda). 
\end{multline}
Here $K(x)$ is a kernel, which, for integer $n$, is the Jacobian of
the $n$-dimensional change of variables from Cartesian to spherical
coordinates. It is still undetermined for non-integer
dimensions. 

\section{Identification of terms dominating convergence}
In the main text, we argued that it is justifiable to drop the
$y^{4-n}$ term in Eq.~(\ref{eqn:al5diff}), or equivalently,
the $y^{5-2n}$ term in Eq.~(\ref{eqn:normdiff}).
In essence, the bulk of the solution is supported
away from $y=0$, while this term is quickly suppressed for
$n>3$. Without this (generally) non-integer power, we were able to
calculate exactly predictive information for our model. Whatever
the contributions the full solution might add, they are of lower order
than the leading term in Eq.~(\ref{eqn:infofinal}). Nonetheless, this
term is crucial since it keeps the full solution physical by
guaranteeing its convergence faster than any power as $y\to 0$
($x\to\infty$ in the $x$ space). In this appendix, we will make the
arguments a bit more precise.

Our approach is of the maximum principle type, which is employed
abundantly in the theory of partial differential equations. We present
the arguments in a general setting, not limited to the confines of our
model.  Our focus is on equations of the type
\begin{equation}
\label{eqn:type}
\partial_t F(t,y)=-g(y)\partial_y F(t,y) + \partial_y^2 F(t,y),\ y>0.
\end{equation}
$F$ is the cumulative probability $\int_0^y f(t,y')\,d y'$ of a
distribution $f$ satisfying a Fokker-Planck equation with constant
noise and a force $g(y)$. We will assume that around $y\thicksim 0$,
$g$ is positive and behaves as $1/y^\alpha$ with $\alpha>1$.  We start
by providing a sort of a zero value ``eigenvector'', i.e. a solution of
the equation
\begin{equation}
\label{eqn:eigen}
0=-g(y)\partial_y F_0(y) + \partial_y^2 F_0(y).
\end{equation}
It is straightforward to see that Eq.~(\ref{eqn:eigen}) is
solved by 
\begin{equation}
\label{eqn:eigen1}
F_0(y)=\int_0^y\,d y'\exp\left(\int_{y_0}^{y'}\,d y''g(y'')\right),
\end{equation}
where $y_0$ is any positive value. It follows that $F_0(y)\thicksim
\exp(-1/y^{\alpha-1})$, thus it converges to zero, together with all
of its derivatives.

The solution, Eq.~(\ref{eqn:eigen}), is non-normalizable, and it is,
therefore, not a true eigenvector. However, we can use it to bound
normalizable solutions of Eq.~(\ref{eqn:type}). That is, we will show
that if initial conditions are bounded everywhere by a multiple of
$F_0$ (e.~g., if their support does not include $0$), then the
solution $F(t,y)$ remains bounded for all times, and it will,
therefore, have all derivatives zero at $y=0$.  This implies that the
exact solution of Eq.~(\ref{eqn:FPgen}) indeed has a finite tail, and
this is all due to the third term in Eq.~(\ref{eqn:normdiff}).  By
imposing the requirement that this term diverges faster than $1/y$, we
obtain $n>3$, or equivalently $\alpha<4$.

In order to demonstrate that $F(t,y)\le F_0(y)$ if $F(0,y)\le F_0(y)$,
we will first show the following.

{\it If $\tilde F(t,y)$ satisfies the boundary conditions $\tilde
  F(t,0)=0$ and $\tilde F(t,L)\ge 1/2$, for some $L>0$, together with
  the initial condition $\tilde F(0,y)\ge 0$, then $\tilde F$ remains
  non-negative for all times if it also satisfies the following
  equation: }
\begin{equation}
\label{eqn:type1}
\partial_t \tilde F(t,y)=-\gamma \tilde F(t,y)-g(y)\partial_y \tilde F(t,y) + \partial_y^2 \tilde F(t,y),\ \gamma>0.
\end{equation}

\begin{proof}
Assume a negative minimum of $\tilde F(t_0,y_0)<-\epsilon$ at some time $t_0$ and
point $y_0$. Clearly, $0<y_0<L$. Then, at $y_0$:
\begin{multline}
\label{eqn:lemma}
\partial_t \tilde F(t_0,x_0)=\partial^2_y \tilde F(t_0,x_0) -g(y_0)\partial_y \tilde F(t_0,y_0) 
\\ - \gamma \tilde F(t_0,y_0) \ge \partial^2_y \tilde F(t_0,x_0) +\gamma\epsilon > 0.
\end{multline}
This implies that there is a $\delta>0$ such that $\tilde F <-\epsilon$ at 
some points $y$, for all $t_0-\delta<t<t_0$. Let $\tilde t$ be the infimum of
the set of all times for which $\tilde F <-\epsilon$ at some point. Take
a sequence $\{t_n\}$ which converges to $\tilde t$ and a sequence $\{y_n\}$ such that
$\tilde F (t_n,y_n) <-\epsilon$. Since $0<\{y_n\}<L$, we can assume that it converges 
to some $\tilde y \ne 0$. Thus, $\tilde F(\tilde t, \tilde y)<-\epsilon$. By applying
Eq.~(\ref{eqn:lemma}) again, we obtain that this is possible only if $\tilde t=0$, which,
in turn, is impossible because of the initial conditions.
\end{proof}

Notice that the positivity of $\tilde F$ immediately implies the
positivity of $F$ since there is a one-to-one mapping between the
solutions of Eqs.~(\ref{eqn:type}) and (\ref{eqn:type1}) given by
$\tilde F\exp(\gamma t)=F$.  If we apply this to $\Delta F(t,y)\equiv
F_0(t,y)-F(t,y)$, then $F_0(t,y)\ge F(t,y)$ for all times $t$, as long
as this is true for $t=0$, just as we claimed earlier. We end with a
comment regarding the boundary condition requirement at $y=L$. If
$F_0$ is non-normalizable, then this condition is trivially
satisfied. Otherwise, this condition is a byproduct of the uniqueness
requirements of the solution.  Therefore, the approximate solution,
Eq.~(\ref{eqn:ugrn}), is an upper bound on the exact solution of
Eq.~(\ref{eqn:FPgen}).

\section{Bounding subleading terms in predictive information}
While we have not been able to obtain a closed form expression for all
terms in Eq.~(\ref{eqn:geninfocont1}), we can nonetheless provide
asymptotically finite bounds on them. We will rely on the basic
structure of the solution, Eq.~(\ref{eqn:fgrn1}), and repeated
applications of the Jensen's inequality.

Starting with the full expression in Eq.~(\ref{eqn:geninfocont1}), we
would like to start by providing the following bounds for $z>0$ and
$\theta>1$, $\vartheta>0$:
\begin{equation}
\label{eqn:bounds}
A(\theta,\vartheta)+B(\theta,\vartheta)z^\theta\ge \int_0^\infty y^\theta 
(y-z)^\vartheta e^{-(y-z)^2/2}\ge a(\theta,\vartheta)z^{\theta-1}
+b(\theta,\vartheta)z^\theta.
\end{equation}
Here $A,\ B,\ a,\ b$ are positive functions of $\theta$ and
$\vartheta$ only.  It is useful to normalize the kernel
$K(\lambda)$. Thus we define
\begin{equation}
\kappa =\int_{-1}^1 K(\lambda) \,d\lambda=2^{n-2}\frac{\Gamma\left(\frac{n-1}{2}\right)^2}
{\Gamma(n-1)},
\end{equation} 
where the last equality contains the usual Gamma function.  We now can
provide an upper bound on the integral terms in
Eq.~\ref{eqn:geninfocont1}. By using the fact that $x\log(x)$ is a
convex function, we obtain
\begin{equation}
\label{eqn:uppbound}
\begin{split}
& C^{-1}(2\pi)^{n/2}\left\langle\log_2 \int_{-1}^1 \,d\lambda K(\lambda)
\Xi (T;\lambda,y,w)\right\rangle - C^{-1}(2\pi)^{n/2}\log_2\kappa =\\ &
\kappa\int_0^\infty\,dy\frac{y^{n-1}}{\Delta^n(T)}\int_{-1}^1 \,d\lambda
\frac{K(\lambda)}{\kappa}\Xi(T;\lambda,y,w)\log_2\int_{-1}^1\,d\lambda' 
\frac{K(\lambda')}{\kappa}\Xi(T;\lambda',y,w) \\ &
\le \kappa\int_0^\infty \,dy \frac{y^{n-1}}{\Delta^n(T)}\,d\lambda\frac{K(\lambda)}{\kappa}
\Xi(T;\lambda,y,w)\log_2\Xi(T;\lambda,y,w) \le \\ &
 \le -(1/2)\log_2(e)\int_{-1}^1 \,d\lambda K(\lambda)\left[a(n-1,2)\lambda^{n-2} 
\left(\frac{we^{\hat\tau T}}{\Delta(T)}\right)^{n-2}+\right. \\ &\quad  b(n-1,2)\lambda^{n-1}
 \left(\frac{we^{\hat\tau T}}{\Delta(T)}\right)^{n-1}+ 
a(n-1,0)(1-\lambda^2)\lambda^{n-2}\left(\frac{we^{\hat\tau T}}{\Delta(T)}\right)^n+ \\ &\quad
 \left. b(n-1,0)(1-\lambda^2)\lambda^{n-1}\left(\frac{we^{\hat\tau T}}{\Delta(T)}\right)^{n+1}
\right]e^{-\frac{1-\lambda^2}{2\Delta(T)^2}w^2e^{2\hat\tau T}}.
\end{split}
\end{equation}
Similarly, utilizing the concavity of $\log(x)$, we can write 
a lower bound on the expectation value
\begin{equation}
\label{eqn:lowbound}
\begin{split}
& C^{-1}(2\pi)^{n/2}\left\langle\log_2 \int_{-1}^1 \,d\lambda K(\lambda) \Xi (T;\lambda,y,w)
\right\rangle  - C^{-1}(2\pi)^{n/2}\log_2\kappa \\ & 
\ge-(1/2)\log_2(e)\int_{-1}^1 \,d\lambda K(\lambda)
\left[A(n+1,0)+B(n+1,0)\lambda^{n+1}\left(\frac{we^{\hat\tau T}}{\Delta(T)}\right)^{n+1} \right. \\
& \quad\left. A(n-1,0)\left(\frac{we^{\hat\tau T}}{\Delta(T)}\right)^{2}+
B(n-1,0)\lambda^{n-1}\left(\frac{we^{\hat\tau T}}{\Delta(T)}\right)^{n+1}\right]
e^{-\frac{1-\lambda^2}{2\Delta^2(T)}w^2e^{2\hat\tau T}}.
\end{split}
\end{equation}
Therefore, we have obtained bounds on the third term in
Eq.~(\ref{eqn:geninfocont1}) that are polynomial in $e^{\hat\tau
  T}/\Delta(T)$.  The latter is, in turn, a bounded function of
$T=t+\tilde t$.  Indeed, it is straightforward to show that
$e^{\hat\tau T}/\Delta(T)\le \sqrt{|\hat\tau|}+\sqrt{1/T}$. Therefore,
these bounds are asymptotically constant (as $T\to \infty$) and either
$\mathcal{O}(1)$ or $\mathcal{O}(\sqrt{|\hat\tau|})$.  We can use
these bounds on the second term of Eq.~(\ref{eqn:geninfocont1}) by
simply replacing $w$ by $z$ and $T$ by $\tilde t$ in
Eqs.~(\ref{eqn:uppbound}) and (\ref{eqn:lowbound}). The resulting
expressions need to be averaged over $z$, which requires estimating
quantities of the form
\begin{multline}
\label{eqn:secondavg}
L\le \int_0^\infty \,dz \frac{z^{n-1}}{\Delta^n(t)}z^m
\left(\frac{e^{\hat\tau \tilde t}}{\Delta(\tilde
    t)}\right)^{m}
e^{-\frac{1-\lambda^2}{2\Delta^2(\tilde t)}e^{2\hat\tau \tilde t}z^2} \\
\times \int_{-1}^1\,d\tilde\lambda
K(\tilde\lambda)\exp\left(-\frac{1}{2\Delta(t)^2}
  (z^2-2zw\tilde\lambda e^{\hat\tau t} + w^2 e^{2\hat\tau t})\right) \le U,
\end{multline}
where $m$ is a positive number. By using Eq.~(\ref{eqn:bounds}) again,
we can obtain an upper and a lower bound on this expression. It is
convenient to introduce $\eta^2=(1-\lambda^2)e^{2\hat\tau \tilde
  t}\Delta^2(t)/\Delta^2(\tilde t)$.  Then, after some algebra, we
obtain the following two bounds: an upper bound
\begin{equation}
\label{eqn:secondupp}
\begin{split}
  & U  = \int_{-1}^1 \,d\tilde\lambda
  K(\tilde\lambda)\left[\frac{\eta^2}{1-\lambda^2}\right]^{m/2}
  (1+\eta^2)^{-(n+m)/2}\left[\vphantom{\left( \frac{\tilde\lambda
          we^{\hat\tau t}}{\Delta(t)(1+\eta^2)^{1/2}}\right)^{n+m-1}}
    A(n+m-1,0)+ \right. \\ & \quad\left.+B(n+m-1,0)\left(
      \frac{\tilde\lambda we^{\hat\tau
          t}}{\Delta(t)(1+\eta^2)^{1/2}}\right)^{n+m-1}\right]
  \exp\left(-\frac{1}{2}\left(1-\frac{\tilde\lambda^2}{1+\eta^2}\right)
    \frac{w^2e^{2\hat\tau t}} {\Delta^2(t)} \right),
\end{split}
\end{equation}
and a lower bound
\begin{equation}
\label{eqn:secondlow}
\begin{split}
  & L = \int_{-1}^1 \,d\tilde\lambda
  K(\tilde\lambda)\left[\frac{\eta^2}{1-\lambda^2}\right]^{m/2}
  (1+\eta^2)^{-(n+m)/2}\left(\frac{\tilde\lambda we^{\hat\tau
        t}}{\Delta(t)(1+\eta^2)^{1/2}}\right) ^{n+m-2} \\ &
  \quad\times\left[a(n+m-1,0)+b(n+m-1,0)\left( \frac{\tilde\lambda
        we^{\hat\tau t}}{\Delta(t)(1+\eta^2)^{1/2}}\right)\right] \\
  &\quad
  \times\exp\left(-\frac{1}{2}\left(1-\frac{\tilde\lambda^2}{1+\eta^2}\right)
    \frac{w^2e^{2\hat\tau t}} {\Delta^2(t)} \right).
\end{split}
\end{equation}
Notice that, for $\hat\tau\ge 0$, $\eta\to\infty$ as $t\to\infty$,
while both bounds in Eqs.~(\ref{eqn:secondupp}) and
(\ref{eqn:secondlow}) are of order $\mathcal{O}(\eta^{-m/2})$,
therefore they are asymptotically constant. For $\hat\tau<0$,
Eqs.~(\ref{eqn:secondupp}) and (\ref{eqn:secondlow}) are controlled by
$\mathcal{O}(|\hat\tau|^{m/2})$. This implies that the second term in
Eq.~(\ref{eqn:geninfocont1}) is also bounded around the critical
point, independently of $\hat\tau$.  This completes the proof that the
terms we dropped in Eq.~(\ref{eqn:geninfocont1}) do not contribute to
the leading order of predictive information.

\bibliographystyle{unsrt}

\end{document}